\documentstyle[prd,aps,floats,epsf]{revtex}
\input epsf.tex
\def\be{\begin{equation}}
\def\ee{\end{equation}}
\def\bea{\begin{eqnarray}}
\def\eea{\end{eqnarray}}

\begin{document}
\title{Quintessence and Spontaneous Leptogenesis}
\author
{Mingzhe Li,  Xiulian Wang,  Bo Feng  and Xinmin Zhang}
\address{Institute of High Energy Physics, \\Chinese
Academy of Science, PO Box 918-4, Beijing 100039,\\ P. R. China}
\maketitle

\begin{abstract}
{We propose in this paper a scenario of spontaneous baryogenesis in
cosmological models of Quintessence by introducing a
derivative coupling of the Quintessence
scalar $Q$ to the baryon current $J_B^{\mu}$ or the current of the baryon
number minus lepton number $J_{B-L}^{\mu}$. We find
that with a dimension-5 operator ${\partial_\mu Q} J_{B-L}^{\mu}$
suppressed by the Planck mass $M_{pl}$ or the Grand Unification Scale
$M_{GUT}$, baryon number asymmetry $n_B/s \sim
10^{-10}$ can be naturally explained {\it via} leptogenesis.
 We study also the isocurvature
baryon number fluctuation generated in our model.}
\end{abstract}


\vskip 0.4cm
%\tighten
%\narrowtext
%\section{Introduction}

Evidence is increasing that Universe is spatially flat and accelerating
at present time\cite{pel}. The simplest account of the cosmic acceleration seems to
be a remnant small cosmological constant $\Lambda$, however many
physicists are attracted by the idea that a new form of matter is causing
the cosmic acceleration. This {\it new form} of matter is not clustered
gravitationally on the scale of the galaxy clusters and has been
usually called {\bf dark energy} or {\bf Quintessence}\cite{rp,we,swz,turner}.

For Quintessence, typically one assumes the existence of a scalar field
$Q$ (or multi scalar fields) which is taken to be homogeneous in space and
via its kinetic and potential energy contributions to the energy-momentum
tensor tuned to provide an equation of state leading to accelerated
expansion, beginning to dominate the matter content of the Universe today.
Models of Quintessence can have interesting cosmological properties
including {\it tracking} behavior.

Generally minimal or non-minimal couplings of Quintessence with
gravitation are considered in the literature\cite{quin}. Being a dynamical field,
Quintessence is expected to have interactions with
the ordinary matters\cite{peccei}, however as argued in Refs.\cite{ca} for most of
cases the couplings are strongly constrained. But there are
exceptions. For example, Carroll\cite{ca} has considered an
interaction of form $Q F_{\mu\nu}{\tilde F^{\mu\nu}}$
with $F_{\mu\nu}$ being the electromagnetic field strength tensor and
studied its implication on the rotation of the plane
of polarization of light coming from distant sources.
In this paper we introduce a
type of interaction of Quintessence with the matter,  which
 in terms of an effective lagrangian is given by
\begin{eqnarray}\label{lagr}
{\cal L}_{eff}=\frac{c}{M }{\partial_{\mu}Q } ~J^{\mu},
\end{eqnarray}
where $M$ is the cut-off scale which for example could be the Planck
mass $M_{pl}$ or the scale of Grand Unification Theory $M_{GUT}$,
and $c$ is the coupling constant which characterizes
the strength of Quintessence interacting with the ordinary matter in
the Standard Model of the electroweak theory.
Specifically we take in this paper $J^{\mu}$ to be the baryon current
$J^{\mu}_B$ or the current of baryon number minus lepton number $J_{B-L}^{\mu}$,
and study their implications on baryogenesis. The lagrangian in Eq.(\ref{lagr})
involves derivative and
obeys the symmetry $Q \rightarrow Q + {\it constant}$ \cite{ca}, so
the Quintessence potential will not be modified by the quantum corrections.

The mechanism of generating the baryon number asymmetry in this paper
follows closely the
spontaneous baryogenesis\cite{ck,tck}.  The term in Eq.(\ref{lagr}), when $\dot{Q}$
is non-zero during the evolution of spatial flat Friedmann-Robertson-Walker Universe,
violates CPT invariance and
generates an effective chemical potential $\mu_b$ for baryons, ${\it i.e.}$,
\begin{eqnarray}
& &{c\over{M }}\partial_{\mu}Q
J^{\mu}_B \rightarrow ~c ~
{\dot{Q}\over{M }}n_{B}= ~c ~{\dot{Q}\over{M }}(n_{b}-n_{\bar
b})~,\nonumber\\
& &\mu_{b}=~c ~~{\dot{Q}\over{M }}=-\mu_{\bar b}~.
\end{eqnarray}
In thermal equilibrium the baryon number asymmetry is given by (when
$T\gg m_{b}$)
\begin{eqnarray}
n_{B}=\frac{g_{b}T^{3}}{6}({\mu_{b}\over T}+{\cal O}({{\mu_{b}\over
T})}^3)\simeq c \frac{g_{b}\dot Q T^{2}}{6 M }~,
\end{eqnarray}
where $g_b$ counts the internal degree of freedom of the baryon.
Using the familiar expression for entropy density
\begin{equation}
s =\frac{2 \pi^2}{45} g_{\star} T^3,
\end{equation}
we arrive at the final expression for the baryon to entropy ratio
\begin{equation}\label{bnumber}
n_{B}/s\simeq \frac{15 c}{4 \pi^2}\frac{g_{b}\dot Q}{g_{\star}M  T}~.
\end{equation}
$\dot Q$ in Eq.(\ref{bnumber}) can be obtained by solving
 the equation of motion of
Quintessence given below
\begin{equation}\label{motion}
\ddot Q+3H\dot Q+V^{'}(Q)=-{ c\over M }(\dot n_{B}+3Hn_{B})~,
\end{equation}
where H is the Hubble constant and $V(Q)$ is the potential of
Quintessence field.
For the radiation dominated era the Hubble constant is
\begin{equation}
H={1\over 2t}=1.66 g^{1/2}_{\star}\frac{T^2}{M_{pl}}~.
\end{equation}
The right-handed side of Eq.(\ref{motion}) is about
$-\frac{c g_b} {6} \frac{T^2 }{M^2 }(\ddot Q+H\dot Q)$
and can be neglected unless in the
very early universe when the temperature $T$ is close to the cut-off scale.

In this paper we consider a model which has the tracking property, where
the potential has a modified exponential form\cite{as},
\begin{equation}\label{potential}
V(Q)=f(Q)e^{-{\lambda\over m_{pl}}Q}~.
\end{equation}
In Eq.(\ref{potential}), $f(Q)$ is a function of Q which is required to
change more slowly than $e^{-{\lambda\over m_{pl}}Q}$
in the regime of tracking. For the detailed discussions on the specific
form of $f(Q)$ and model properties we refer to Ref.\cite{as}.
Here we simply summarize the salient features of this model.
For $\lambda^2>3(1+w_{B})$ with $w_B$ being the ratio of the pressure to
the energy density of the background fluid, this model has an
attractor solution\cite{hb,ls,fj,bin},
\begin{eqnarray}\label{solution}
\Omega_{Q}&\equiv
&\frac{\rho_{Q}}{\rho_{Q}+\rho_{B}}={3\over\lambda^2}(1+w_{B})~,\nonumber\\
w_{Q}&=&w_{B}~.
\end{eqnarray}

During radiation dominated era, $\rho_B = \frac{\pi^2}{30} g_{\star}
T^4$ and $w_B = 1/3$, so we have
\begin{eqnarray}\label{track}
\rho_{Q}&\simeq& \frac{2 \pi^2}{ 15 (\lambda^2 - 4)}
g_{\star}T^4~,\nonumber\\
\dot Q&\simeq& \frac{4\pi}{3\sqrt{10(\lambda^2-4)}}
g_{\star}^{1/2} T^2~.
\end{eqnarray}
Thus we obtain the ratio of baryon number to entropy, which is given by
\begin{eqnarray}
 {n_{B}\over s}\simeq
\frac{5c}{\pi\sqrt{10(\lambda^2 -4 )}}g_b
g_{\star}^{-1/2} \frac{T }{ M }~.
\end{eqnarray}
Taking  $\lambda^2 >20$ \cite{fj,bin} constrained by
nucleosynthesis, and $g_{\star} \sim {\cal O}(100), ~~ g_b \sim
{\cal O}(1)$ we have
\begin{eqnarray}
 {n_{B}\over s}\sim 0.01 ~ c ~~ \frac{T}{M }.
\end{eqnarray}

In the scenario of spontaneous baryogenesis, the baryon asymmetry is
generated in thermal equilibrium. This requires baryon number violating
interactions occur rapidly ($\Gamma_{\not B} > H$). However if the B-violating
interactions keep in equilibrium until $\dot Q \rightarrow 0$, the final
baryon asymmetry will be zero. Denoting the epoch
when the B-violating interactions
freeze out by $T_D$, {\it i.e.} $\Gamma_{\not B} (T_D) \simeq H( T_D )$, the
final baryon number asymmetry obtained is
\begin{eqnarray}\label{fnumber}
 {n_{B}\over s}|_{T_{D}}\sim 0.01 ~ c ~~ \frac{T_D}{M }.
\end{eqnarray}

For $T_D$ around the Grand Unification scale $10^{16}GeV$,
Eq.(\ref{fnumber}) shows that
it is quite easy to have
$n_B / s \sim 10^{-10}$ required by
nucleosynthesis and measurement of Cosmic Microwave Background
anisotropy, for both $M= M_{pl}$ and $M=M_{GUT}$.
However, if the B-violating interactions of Grand Unified Theory
conserve $B-L$, the asymmetry generated will be erased by the
electroweak Sphaleron \cite{manton}. In this case Sphaleron $B+L$
interactions will make $T_{D}$ as low as around 100 GeV and
$n_{B} / s$ generated will be too small unless $c$ is unreasonably large.

Now we discuss the mechanism of leptogenesis \cite{zhang}
in our model by identifying $ J^{\mu}$ with $J_{B-L}^\mu$
in Eq.(\ref{lagr}).
Doing the calculation with the same procedure as above for $J^{\mu} =
J^{\mu}_{B}$
we have the final asymmetry of the baryon number minus lepton number
\begin{eqnarray}\label{fnumber2}
 {n_{B-L}\over s}\sim 0.01 ~ c ~~ \frac{T_D}{M }.
\end{eqnarray}
The asymmetry $n_{B-L}$ in (\ref{fnumber2})  will be converted to baryon number
asymmetry when electroweak Sphaleron $B+L$
interaction is in thermal equilibrium which happens for
temperature in the range of $10^2 {\rm GeV}
\sim 10^{12}{\rm GeV}$. $T_D$ in (\ref{fnumber2}) is the temperature below which
the $B-L$ violating interactions freeze out.

In the Standard Model of electroweak theory, $B-L$ symmetry is exactly
conserved, however it could be violated by the (majorana) neutrino masses
{\it via} higher dimensional effective operators. For instance, we take
a dimension-5 operator,
\begin{eqnarray}\label{lepvio}
    {\cal L}_{\not L} = \frac{2} { f } l_L l_L \phi \phi ~~ + ~~ {\it
h.c.},
\end{eqnarray}
where $f$ is a scale of new physics beyond the Standard Model which
generates the $B-L$ violations, $l_{L}$
and $\phi$ are the left-handed lepton and Higgs doublets respectively.
When the Higgs
field gets vacuum expectation value $< \phi > \sim v $, the
left-handed neutrino receives a majorana mass $m_\nu \sim
\frac{v^2}{f}$.

In the early universe the lepton number violating rate induced by the
interaction in (\ref{lepvio}) is\cite{sarkar}
\begin{eqnarray}
  \Gamma_{\not L} \sim
    0.04~ \frac{T^3}{ f^2 }.
\end{eqnarray}
This rate at $T_D$
is required to be slower than the Universe expansion rate
$\sim 1.66 g_{\star}^{1/2}T^2/ M_{pl}$, which will put a lower limit on
the
scale $f$, consequently a upper limit on the neutrino mass $m_\nu$
\cite{barr}.

To get the upper limit on $m_\nu$, we try to minimize the value of $f$.
This can be done by minimizing the freezing out temperature
$T_D$. From Eq.(14), one can see that if requiring $c \leq 4 \pi$ so that the
effective lagrangian (Eq.(1)) is sensible for the perturbative calculation,
$T_D$ must be larger than $10^{10}$ GeV for $M=M_{pl}$, and $10^7$GeV
for $M = M_{GUT} \sim 10^{16}$GeV. As a result of it,
$f$ should be larger than about $10^{13}$ GeV and the neutrino
mass should be smaller than $\sim 4$ eV for $M=M_{pl}$. In the case that
$M=M_{GUT}, ~ f > 4.9 \times 10^{11}$GeV and
$m_\nu < 0.1$keV. And for both cases, $T_D$ are much smaller than the
cut-off scales. This makes our approximation of
neglecting the {\it r.h.s} of Eq.(6) valid. We have also done
a numerical calculation with result which supports
for it.

So far what we have considered is the case that Quintessence is homogeneous in space. Since
Quintessence is a scalar field one expects its fluctuations in space-time.
In the following we study the isocurvature baryon number fluctuations
generated in our model.

Since the baryon number asymmetry is proportional to $\dot Q$ and
in the tracking regime $\dot Q\propto \sqrt{V(Q)}$, we have
\begin{eqnarray}
\frac{n_{B}}{s}\propto \sqrt {V(Q)},
\end{eqnarray}
and
\begin{eqnarray}\label{dbdq}
\delta(\frac{n_{B}}{s})\propto \frac{V^{'}}{2\sqrt{V}}\delta
Q~,\nonumber\\
\frac{\delta(n_{B}/s)}{n_{B}/s}=\frac{V^{'}}{2V}\delta Q~.
\end{eqnarray}
To get $\delta Q$, we are solving the equation of
motion given below,
\begin{eqnarray}\label{eqperturbation}
\ddot{\delta Q_{k}}+3H\dot{\delta Q_{k}}+\frac{k^2}{a^2}\delta Q_{k}+V{''}\delta
Q_{k}=0~,
\end{eqnarray}
where $\delta Q_{k}$ is the Fourier transform of $\delta Q$.
To have an analytical formula for $\delta Q$ we closely follow
\cite{bmr,af,kmt} by writing $V^{''}$ as
\begin{eqnarray}\label{vpp}
V^{''}=\frac{3}{2}H{\dot c_{Q}^2}+ \frac{9}{4}(1-c_{Q}^2)
(w_{B}+c_{Q}^2+2)H^2,
\end{eqnarray}
where $c_{Q}^2$ is the sound speed of Quintessence defined by
\begin{equation}
c_{Q}^2\equiv \frac{\dot p_{Q}}{\dot \rho_{Q}}=w_{Q}-\frac{\dot
w_{Q}}{3H(1+w_{Q})}~.
\end{equation}
Consider the case that the sound speed
is approximately constant\cite{af}, the
first term of the {\it rhs.} of
Eq.(\ref{vpp}) can be dropped out, then the solution to
Eq.(\ref{eqperturbation}) is\cite{af}:
\begin{equation}\label{solution2}
\delta Q_{k}\simeq \eta^{-1/2}[C_{1}J_{\nu}(k\eta)+C_{2}J_{-\nu}(k\eta)]~,
\end{equation}
where $\eta$ is the conformal time ($d\eta\equiv \frac{dt}{a}$),
and $a\propto \eta \propto T^{-1}$ during the radiation dominated era.
In Eq.(\ref{solution2}),
$J_{\pm\nu}(k\eta)$ are the first kind of Bessel
functions, $C_{1}$, $C_{2}$ are constants which are time
independent, $\nu^2$ and $\beta$ are,
\begin{eqnarray}
& &\nu^2\equiv {1\over 4}-\beta~,\nonumber\\
& &\beta\equiv\frac{9}{4}(1-c_{Q}^2)(w_{B}+c_{Q}^2+2)~.
\end{eqnarray}

The modes of fluctuations which are interesting in cosmology are
those outside the Hubble radius in very
early universe, {\it i.e.} $k\eta\ll 1$. In such a limit the Bessel
function can be approximated by
\begin{equation}
J_{\nu}(k\eta)\simeq \frac{(k\eta)^{\nu}}{2^{\nu}\Gamma (1+\nu)}~, (k\eta\rightarrow 0)~.
\end{equation}

On the initial value of $\delta Q_k$ denoted by $\delta Q_{ki}$, we assume
the primordial fluctuation of the Quintessence scalar is
generated during inflation. Being almost massless $\delta Q_{ki}$ is simply
\begin{equation}\label{dqi}
|\delta Q_{ki}|=\frac{H_{in}}{\sqrt{2k^3}}~,
\end{equation}
where $H_{in}$ is the Hubble constant during inflation.
Furthermore the kinetic energy of the Quintessence will be diluted by
inflation. Thus the Quintessence energy density remains potential energy
dominated after inflation until the time $\eta_t$ when it goes into the tracking regime.
This period of time is called the {\it potential phase} in the Ref.\cite{af}.
During this period $\rho_{Q}$ is nearly constant and
$c_{Q}^2\simeq -w_{B}-2$ \cite{bmr}, hence $\beta\simeq
0$ and $V^{''}\simeq 0$. We can see that $\delta Q_{k}$ is constant in
the potential phase. But in the regime of tracking ($\eta \geq \eta_{t}$),
\begin{eqnarray}
c_{Q}^2&=&w_{Q}~,\nonumber\\
\beta&=&\frac{9}{4}(1-w_{Q})(w_{B}+w_{Q}+2)=\frac{9}{4}(1-w_{Q})({7\over
3}+w_{Q})~.
\end{eqnarray}
Numerically $-1<w_{Q}\leq 1/3$, so we have
$\beta>1/4$. We obtain that
\begin{eqnarray}\label{jnu}
& &\nu^2<0~,\nonumber\\
& &\nu=i|\nu|=i\sqrt{\beta-{1\over 4}}~\nonumber\\
& &J_{\pm \nu}(k\eta)\simeq\frac{\exp(\pm i|\nu|\ln{k\eta\over
2})}{\Gamma(1\pm i|\nu|)}~.
\end{eqnarray}
Substituting $J_{\pm\nu}(k\eta)$ above and $\delta Q_{ki}$ (Eq.(\ref{dqi})) into (\ref{solution2}),
we will get $\delta Q_k$.

We have also numerically solved Eq.(\ref{eqperturbation}) for the Quintessence model we use
in this paper. In Fig.1 we plot $\delta Q_k$ as a function of red-shift
$z$. One can see that $\delta Q_{k}$ is constant in the potential phase {\it i.e.}
$\delta Q_{k i} = \delta Q_{k t}$, then oscillates with
amplitude which decays like $\propto \eta^{-1/2}$ during tracking.
This confirms our analytical analysis above.

Denoting the $\delta Q_k$ at the
temperature $T_D$ when the baryon number asymmetry freezes in by
$\delta Q_{kD}$, we have the ratio of $|\delta Q_{kD}|$ to the
$|\delta Q_{ki}|$
\begin{eqnarray}
|\frac{\delta Q_{kD}}{\delta Q_{ki}}|\leq ({\eta_{t}\over
\eta_{D}})^{1/2}=({T_{D}\over T_{t}})^{1/2}~,
\end{eqnarray}
where $T_{t}$ is the temperature when $Q$ transits from the potential
phase to the tracking regime.
Making use of the initial value of $\delta Q_{k}$ generated
during inflation, $\delta Q_{ki}=\frac{H_{in}}{\sqrt{2k^3}}$, we have
\bea
|\delta Q_{kD}|&\leq &\frac{H_{in}}{\sqrt{2k^3}}({T_{D}\over
T_{t}})^{1/2}~.
\eea
Following \cite{tck}, we calculate the RMS
fluctuation of baryon number per logarithmic interval in our model which is given by
\be
(\frac{\delta n_{B}}{n_{B}})_{k}\leq \frac{H_{in}}{4\pi}| V^{'}(Q_{D})/V(Q_{D})|({T_{D}\over
T_{t}})^{1/2}~.
\ee
For the Quintessence model with potential in Eq.(\ref{potential}),
\bea\label{result}
(\frac{\delta n_{B}}{n_{B}})_{k}&\lesssim & {\lambda\over 4\pi}\frac{H_{in}}{M_{pl}}(\frac{T_{D}}{T_{t}})^{1/2}\nonumber\\
&\leq &10^{-5}(\frac{T_{D}}{T_{t}})^{1/2}~,
\eea
which will be smaller than $10^{-5}$ since $T_{t} > T_D$.
To get the result above we have used the constraints on
$\frac{H_{in}}{M_{pl}}\leq 10^{-5}$\cite{lr} from the limit of the detecting of primordial gravitational
waves predicted by inflation models. In addition we have also taken $\lambda^2\sim {\cal O}(10)$
to satisfy the constraints from the nucleosynthesis.
The isocurvature baryon number fluctuation in (\ref{result}) is small and
is not conflict with the Boomerang and Maxima-1 data on the anisotropy of
cosmic microwave background\cite{ekv}.

The isocurvature baryon number fluctuation in the spontaneous
baryogenesis\cite{ck} has been studied
by Turner, Cohen and Kaplan \cite{tck}, where they find the amplitude sizable.
In their model the potential of the {\it ilion} field
$\theta$ is $m^2 \theta^2$. The produced baryon
asymmetry is linearly proportional to the initial value of the
ilion field $\theta_{0}$.
Therefore quantum fluctuations in the initial
$\theta_{0}$ during inflation will directly induce isocurvature
baryon number fluctuations. In our model the Quintessence has
the tracking behavior, and the produced baryon number asymmetry is
almost independent of the initial conditions of Quintessence
field. Consequently a small isocurvature perturbation is expectable.

The {\it tracking} behavior of the Quintessence, however plays an important
role in our discussions. Because of the tracking,
$\dot Q$ can be large enough and the baryon number
asymmetry required can be generated. For models
without tracking property, such as
the model with potential $V(Q) = \Lambda^4 ( 1 \pm \cos\frac{Q}{M})$, we
have checked and
found that $\dot Q$ is too small to account for $n_B / s \sim 10^{-10}$.

Our mechanism for leptogenesis, however does put a lower limit on the
reheating temperature $T_{R}$. It requires $T_R$ be higher
than $T_{t}$ and $T_D$. For cut-off $M$ of Eq.(1) in the range of
$10^{16} \sim 10^{19}$GeV, $T_D$ is about $10^{7} \sim 10^{10}$GeV, which
will not put a strong constraint on models of inflation and reheating.

In summary we have introduced in this paper an effective
interaction of the Quintessence scalar to the ordinary matter and studied the
possibilities of baryogenesis. Our model can naturally explain the
baryon number asymmetry of our Universe $n_B/s \sim 10^{-10}$
{\it via} spontaneous leptogenesis.
Our scenario provides a unified description for the present accelerated expansion and
baryon number asymmetry generation of our Universe by the dynamics of Quintessence.

{\bf{Acknowledgments:}} We thank David Kaplan for correspondence on
spontaneous baryogenesis
and Robert Brandenberger, Fabio Finelli, Z-H. Lin for discussions.
This work is supported in part by National Natural
Science Foundation of China under Grant No. 10047004 and by Ministry of
Science and Technology of China under Grant No. NKBRSF G19990754.

%\section{Conclusion}

{}

\newpage
\topmargin 0.4cm
\begin{figure*}[thb]
%\centerline{\epsfysize 1.5 truein \epsfbox{3.eps}}
\epsfysize=4in
\epsffile{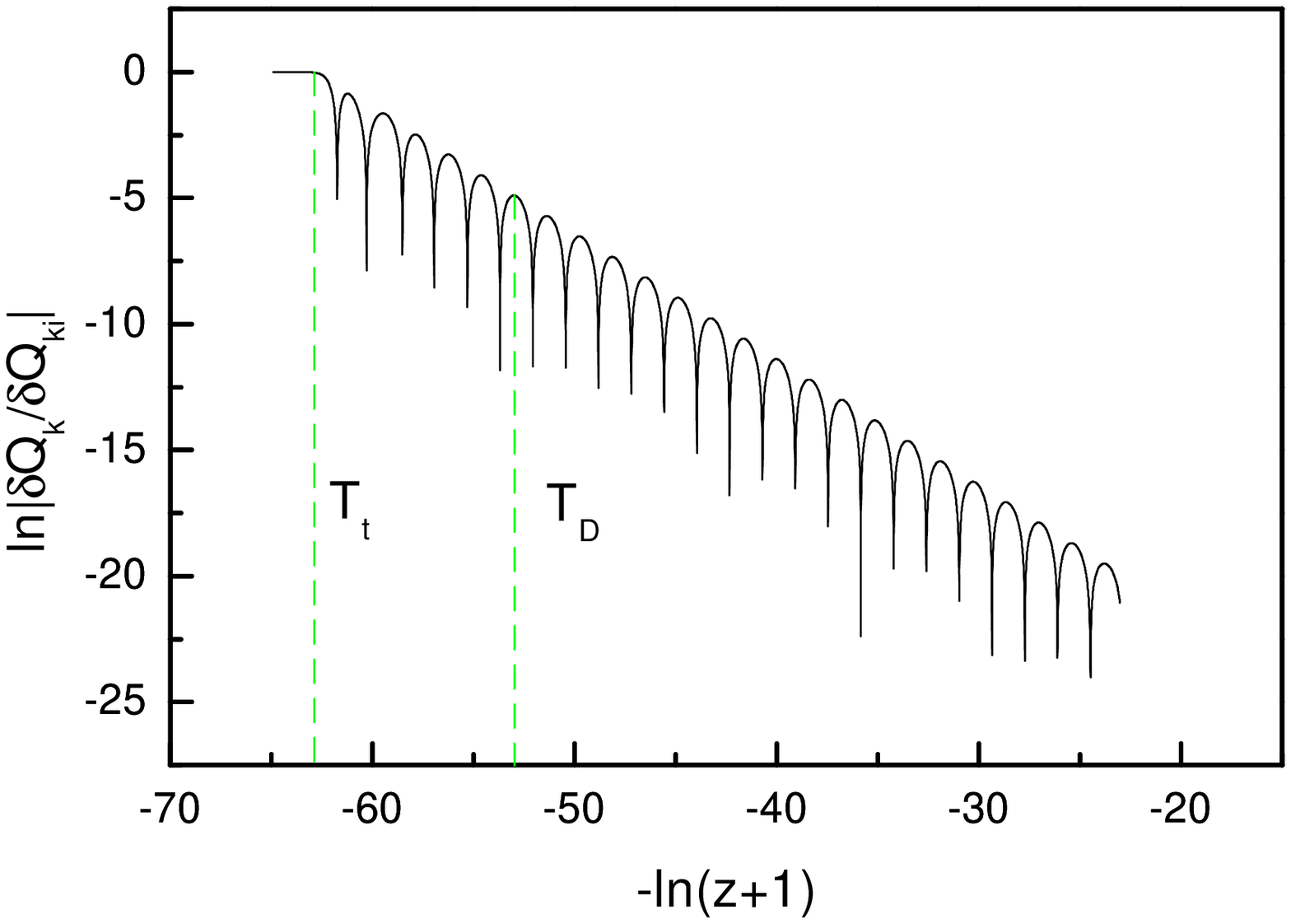}
\vskip 0.3cm
\caption{Plot of $\delta Q_{k}$ as function of red-shift $z$.
The $x$-axis is $- \ln (z+1)$, and the $y$-axis is
$\ln | \delta Q_k / \delta Q_{ki} | $. In this plot, $T_{t}$ corresponds to
temperature around $10^{14}$ GeV and $T_{D} \simeq 10^{10}$ GeV.
}
\end{figure*}

\end{document}